# The model parameters of the mean light curves of the variable red giant stars in the near infrared colour-bands and compare them with the visual mean light curves

**L. S. Kudashkina**

Odessa National Maritime University
kudals04@mail.ru

The observational data of the near infrared bands (H and K) have been used for the modeling mean light curves. Also the visual observational data have been fitted the same.

The infrared and visual mean light curves were compared. All parameters and Fourier-coefficients of the mean light curves were obtained. The periodogram analysis of the variation of the brightness have been carried out.

The observational data of the near infrared bands (H and K) from the article by Whitelock P.A., Marang F., Feast M. «Infrared colours for Mira-like long-period variables found in the Hipparcos catalogue» (2000) were used. 10 stars were processed. Its list in the table 1.

***Table 1. The list of the stars.***

| **Star** | **Number of observations** | **Type** | **Spectral type** | Distance, kpc | Period, d |
|---|---|---|---|---|---|
| **o Cet** | **104** | Mira | M5e-M9e | 0.12 | 331 |
| **R Leo** | **53** | Mira | M6e-M9.5e | 0.11 | 309 |
| **S Car** | **34** | Mira | M2-M3e | 0.51 | 149 |
| **U Her** | **20** | Mira | M6.5e-M9.5e | 0.38 | 406 |
| **X Oph** | **29** | Mira | K1IIIv comp | 0.24 | 328 |
| **R Aql** | **45** | Mira | M5e-M9e | 0.24 | 284 |
| **RR Aql** | **29** | Mira | M6e-M9 | 0.54 | 394 |
| **S Ori** | **101** | Mira | M6.5e-M9.5e | 0.43 | 414 |
| **S Scl** | **122** | Mira | M7-M8IIIe | 0.47 | 362 |
| **$L_2$ Pup** | **77** | SRb | M5e | - | 140 |

The period is from General Catalogue of Variable Stars (GCVS) in this table.

For the analysis we have used the program by Andronov (1994, 2003) which allows the use of a trigonometric polynomial fit:

$$m(t) = a_0 - \sum_{k=1}^{s} r_k \cos(2\pi k \cdot (t - t_k)/P),$$

were $r_k$ are semiamplitudes and $t_k$ are initial epochs for the brightness maximum (minimum magnitude) of the wave with a period $P_k = P/k$.

The preliminary value of the period was corrected by using the method of differential corrections for each order $s$ of the trigonometric polynomial. Next, the r.m.s. residuals from the fit were analyzed using Fischer's criterion, and the value of $s$ corresponding to the statistical significance of the last harmonic (≥ 0.99) was determined.

All computed parameters of light curves are subdivided into three groups: first, fundamental (period P, amplitude $\Delta m = m_{min} - m_{max}$, asymmetry $f = \varphi_{max} - \varphi_{min}$, degree of the trigonometric polynomial *s*); second, parameters of the extremal slope of the light curve; third, additional (parameters of harmonics).

$T_{0max}$ and $T_{0min}$ – epoch extrema; $m_i$ and $m_d$ – the maximal slope of the incline for ascending and descending branches; $t_i$ and $t_d$ – the characteristic time of the increase of brightness by $1^m$ for ascending and descending branches; $\varphi_k$ – is the phase of the maximum of this wave; and $\varphi_m$ is the phase of maximum of the composite fit. Thus $\varphi_3 - 3\varphi_1$ is the phase of maximum of the second harmonic in respect to the main wave; $m_{is}$ and $m_{ds}$ – the ratio of the maximal slope to that obtained for a pure sinusoid of the same period P and amplitude Δm for ascending and descending branches.

The results are given in the tables 2 and 3. For the most stars, the values of the periods are close to GCVS-data. The main light curves are showed in the pictures.

The light curves of almost all stars in H- and K-band are symmetrical, that is the one sinusoidal wave (*s*=1). The exceptions are the light curves of the next stars: R Leo, R Aql in K-band and S Ori, S Scl in both bands (*s*=2). These curves are wide minimum or the hump near minimum. But the error in these phases is sufficiently large because of deficit of the observations in the minimum of the brightness. Contrary, to the similar parameters in the visual region fashion very asymmetrical light curves.

***Table 2.1. The model parameters and the Fourier-coefficients for the mean light curves in the H-band.***

| Star | O Cet | R Aql | R Leo | RR Aql | S Car |
|---|---|---|---|---|---|
| **P, d** | 333.52±.37 | 280.01±.43 | 312.79±.06 | 390.89±1.70 | 149.94±.11 |
| **Δm** | 1.055±.025 | .657±.035 | .677±.032 | 1.199±.047 | .578±.040 |
| **$T_{0\,max}$** | 7895±2 | 6871±3 | 7983±4 | 4209±5 | 6338±3 |
| **$T_{0\,min}$** | 8061±2 | 6731±3 | 7827±4 | 4404±5 | 6263±3 |
| **$m_{max}$** | -2.60±.03 | -.57±.03 | -2.38±.03 | .48±.06 | 1.77±.04 |
| **$m_{min}$** | -1.55±.03 | .09±.04 | -1.70±.03 | 1.68±.04 | 2.35±.04 |
| **s** | 1 | 1 | 1 | 1 | 1 |
| **$m_i$** | -.0099±.0004 | -.0074±.0007 | -.0068±.0005 | -.0096±.0006 | -.0121±.0014 |
| **$m_d$** | .0099±.0004 | .0074±.0007 | .0068±.0005 | .0096±.0006 | .0121±.0014 |
| **$t_i$** | -101±4 | -136±12 | -147±11 | -104±6 | -83±9 |
| **$t_d$** | 101±4 | 136±12 | 147±11 | 104±6 | 83±9 |
| **$m_{is}$** | 1.00±.04 | 1.00±.09 | 1.00±.08 | 1.00±.07 | 1.00±.04 |
| **$m_{ds}$** | 1.00±.04 | 1.00±.09 | 1.00±.08 | 1.00±.07 | 1.00±.04 |
| **$r_1$** | .53±.02 | .33±.03 | .34±.03 | .60±.04 | .29±.03 |
| **$\varphi_{max}$ ($r_1$)** | -.05±.01 | .18±.01 | .03±.01 | -.05±.01 | .05±.02 |

***Table 2.1 (continue). The model parameters and Fourier-coefficients for the mean light curves in the H-band.***

| Star | $L_2$ Pup | U Her | S Ori | S Scl | X Oph |
|---|---|---|---|---|---|
| **P, d** | 137.14±.19 | 407.00±.67 | 414.33±.50 | 367.03±.25 | 332.56±.74 |
| **Δm** | .353±.028 | .959±.029 | .496±.020 | .886±.016 | .466±.022 |
| **$T_{0\,max}$** | 7974±3 | 8320±5 | 7335±3 | 6795±3 | 7465±6 |
| **$T_{0\,min}$** | 7905±3 | 8523±5 | 7086±11 | 6921±6 | 7631±6 |
| **$m_{max}$** | -2.11±.03 | -.20±.04 | .10±.02 | .29±.02 | -.81±.04 |
| **$m_{min}$** | -1.75±.03 | .76±.05 | .59±.02 | 1.17±.06 | -.35±.02 |
| **s** | 1 | 1 | 2 | 2 | 1 |
| $m_i$ | -.0081 ±.0011 | -.0074±.0005 | -.0045±.0004 | -.0093 ±.0005 | -.0044±.0005 |
| $m_d$ | .0081 ±.0011 | .0074±.0005 | .0051±.0004 | .0114 ±.0010 | .0044±.0005 |
| $t_i$ | -124 ±16 | -135±9 | -224±19 | -108 ±6 | -227±24 |
| $t_d$ | 124 ±16 | 135±9 | 197±16 | 87 ±8 | 227±24 |
| $m_{is}$ | 1.00±.13 | 1.00±.07 | 1.19±.10 | 1.22 ±.07 | 1.00±.11 |
| $m_{ds}$ | 1.00±.13 | 1.00±.07 | 1.35±.11 | 1.51 ±.14 | 1.00±.11 |
| $r_1$ | .18±.02 | .48±.03 | .24±.01 | .38 ±.01 | .23±.03 |
| $\varphi_{max}$ ($r_1$) | .31±.02 | -.45±.01 | .19±.01 | .49 ±.02 | -.04±.02 |
| $r_2$ | | | .06 ±.01 | .16 ±.02 | |
| $\varphi_k(r_2) - 2\varphi_k(r_1)$ | | | .44 ±.03 | .07 ±.02 | |

***Table 2.2. The model parameters and the Fourier-coefficients for the mean light curves in the K-band.***

| Star | O Cet | R Aql | R Leo | RR Aql | S Car |
|---|---|---|---|---|---|
| P, d | 333.79±.39 | 280.60±.34 | 311.03±.83 | 390.67±1.74 | 146.77±.13 |
| Δm | .819±.020 | .555±.034 | .503±.029 | .945±.038 | .490±.037 |
| $T_{0\ max}$ | 7897±2 | 6871±4 | 7993±5 | 4210±5 | 6273±3 |
| $T_{0\ min}$ | 8064±2 | 6765±9 | 8121±21 | 4405±5 | 6346±3 |
| $m_{max}$ | -2.95±.02 | -1.04±.03 | -2.87±.03 | .01±.05 | 1.55±.04 |
| $m_{min}$ | -2.13±.02 | -.49±.04 | -2.37±.03 | .95±.03 | 2.04±.04 |
| s | 1 | 2 | 2 | 1 | 1 |
| $m_i$ | -.0077±.0003 | -.0087±.0011 | -.0069±.0008 | -.0076±.0005 | -.0105±.0014 |
| $m_d$ | .0077±.0003 | .0073±.0010 | .0070±.0008 | .0076±.0005 | .0105±.0014 |
| $t_i$ | -130±6 | -115±15 | -144±17 | -132±8 | -95±12 |
| $t_d$ | 130±6 | 137±19 | 144±16 | 132±8 | 95±12 |
| $m_{is}$ | 1.00±.04 | 1.40±.18 | 1.37±.16 | 1.00±.06 | 1.00±.06 |
| $m_{ds}$ | 1.00±.04 | 1.17±.17 | 1.37±.15 | 1.00±.06 | 1.00±.06 |
| $r_1$ | .41±.02 | .26±.02 | .25±.02 | .47±.03 | .25±.03 |
| $\varphi_{max}$ ($r_1$) | -.02±.01 | .20±.01 | .06±.01 | -.04±.01 | -.40± .02 |
| $r_2$ | | .08 ±.02 | .07 ±.02 | | |
| $\varphi_k(r_2) - 2\varphi_k(r_1)$ | | .32 ±.05 | .12 ±.05 | | |

***Table 2.2 (continued). The model parameters and the Fourier-coefficients for the mean light curves in the K-band.***

| Star | $L_2$ Pup | U Her | S Ori | S Scl | X Oph |
|---|---|---|---|---|---|
| P, d | 137.19 ±.18 | 406.87±.74 | 414.00±.50 | 366.95±.28 | 332.80±.76 |
| Δm | .290 ±.022 | .748±.028 | .408±.018 | .702±.015 | .400±.020 |
| $T_{0\ max}$ | 7974 ±3 | 8325±6 | 7341±4 | 6798±4 | 7474±7 |
| $T_{0\ min}$ | 8043 ±3 | 8529±6 | 7171±18 | 6922±8 | 7641±7 |
| $m_{max}$ | -2.48 ±.03 | -.64±.04 | -.32±.02 | -.06±.02 | -1.20±.03 |
| $m_{min}$ | -2.19 ±.02 | .11±.04 | .09±.02 | .64±.06 | -.80±.02 |
| s | 1 | 1 | 2 | 2 | 1 |
| $m_i$ | -.0066±.0009 | -.0058±.0005 | -.0042±.0004 | -.0082±.0005 | -.0038±.0004 |
| $m_d$ | .0066±.0009 | .0058±.0005 | .0041±.0004 | .0092±.0010 | .0038±.0004 |
| $t_i$ | -151±19 | -173±14 | -238±20 | -122±7 | -265±29 |
| $t_d$ | 151±19 | 173±14 | 246±22 | 108±12 | 265±29 |
| $m_{is}$ | 1.00±.13 | 1.00±.08 | 1.36±.11 | 1.36±.08 | 1.00±.10 |
| $m_{ds}$ | 1.00±.13 | 1.00±.08 | 1.31±.12 | 1.54±17 | 1.00±.10 |
| $r_1$ | .14±.02 | .37±.03 | .20±.01 | .30±.01 | .20±.02 |
| $\varphi_{max}$ $(r_1)$ | -.18±.02 | -.44±.02 | .23±.01 | -.49±.02 | -.012±.02 |
| $r_2$ | | | .06 ±.01 | .14 ±.02 | |
| $\varphi_k(r_2) - 2\varphi_k(r_1)$ | | | .44 ±.03 | .07 ±.02 | |

***Table 3.1. The model parameters and the Fourier-coefficients for the visual mean light curves.***

| Star | O Cet | R Aql | R Leo | RR Aql | S Car |
|---|---|---|---|---|---|
| P, d | 333.33±.02 | 280.84±.01 | 314.16±.02 | 391.7±.4 | 150.05±.01 |
| Δm | 5.41±.02 | 4.54±.02 | 4.08±.01 | 5.09±.06 | 2.80±.01 |
| $T_{0\ max}$ | 6500.2±.7 | 6546.4±.7 | 6367.3±.5 | 3754±2 | 6467.2±.8 |
| $T_{0\ min}$ | 6382±2 | 6428.8±.9 | 6230±3 | 3582±6 | 6392.5±.3 |
| $m_{max}$ | 3.64±.01 | 6.48±.01 | 5.81±.01 | 8.92±.05 | 5.91±.01 |
| $m_{min}$ | 9.05±.02 | 11.02±.02 | 9.89±.02 | 14.0±.1 | 8.71±.02 |
| s | 6 | 9 | 8 | 4 | 4 |
| $m_i$ | -.1072±.0011 | -.062±.002 | -.061±.001 | -.086±.003 | -.077±.002 |
| $m_d$ | .0453±.0010 | .046±.002 | .035±.001 | .040±.005 | .062±.002 |
| $t_i$ | -9.3±.1 | -16.0±.6 | -16.5±.3 | -11.6±.4 | -13.0±.3 |
| $t_d$ | 22.1±.5 | 22±1 | 29±1 | 25±3 | 16.0±.4 |
| $m_{is}$ | 2.10±.02 | 1.23±.05 | 1.49±.03 | 2.11±.08 | 1.31±.03 |
| $m_{ds}$ | .89±.02 | .90±.04 | .85±.03 | 1.0±.1 | 1.07±.03 |
| $r_1$ | 2.64±.01 | 2.14±.01 | 1.867±.005 | 2.65±.5 | 1.28±.01 |
| $\varphi_{max}$ $(r_1)$ | .3289±.0004 | .4433±.0004 | .2446±.0004 | .121±.003 | .272±.001 |
| $r_2$ | .63±.01 | .15±.01 | .241±.005 | .42±.06 | .26±.01 |
| $\varphi_k(r_2) - 2\varphi_k(r_1)$ | .491±.002 | -.382±.006 | .380±.003 | .03±.02 | .013±.004 |

***Table 3.2 (continue). The model parameters and the Fourier-coefficients for the visual mean light curves.***

| Star | $L_2$ Pup | U Her | S Ori | S Scl | X Oph |
|---|---|---|---|---|---|
| P, d | 137.16±.02 | 407.06±.02 | 414.46±.06 | 367.86±.04 | 332.37±.04 |
| Δm | .97±.01 | 4.77±.01 | 4.52±.04 | 6.02±.04 | 1.53±.01 |
| $T_{0\ max}$ | 6590.1±.6 | 6623.3±.4 | 6861.5±.7 | 7112.6±.6 | 6733.6±.6 |
| $T_{0\ min}$ | 6654±1 | 6857±1 | 6640±3 | 7308±6 | 6897±2 |
| $m_{max}$ | 4.34±.01 | 7.69±.01 | 8.40±.02 | 6.75±.02 | 7.15±.01 |
| $m_{min}$ | 5.31±.01 | 12.46±.02 | 12.92±.04 | 12.78±.07 | 8.68±.01 |
| s | 2 | 7 | 4 | 5 | 5 |
| $m_i$ | -.0212±.0008 | -.0748±.0009 | -.0366±.0009 | -.066±.001 | -.0143±.0005 |
| $m_d$ | .0252±.0008 | .0286±.0009 | .035±.001 | .046±.002 | .0167±.0005 |
| $t_i$ | -47±2 | -13.4±.2 | -27.3±.7 | -15.2±.3 | -70±2 |
| $t_d$ | 40±1 | 35±1 | 28±1 | 22±1 | 60±2 |
| $m_{is}$ | .95±.04 | 2.03±.03 | 1.07±.03 | 1.28±.03 | .99±.03 |
| $m_{ds}$ | 1.14±.04 | .78±.02 | 1.03±.03 | .90±.05 | 1.15±.03 |
| $r_1$ | .48±.01 | 2.05±.01 | 2.00±.01 | 2.75±.04 | .716±.004 |
| $\varphi_{max}$ $(r_1)$ | -.316±.003 | -.0242±.0005 | .089±.001 | -.376±.002 | -.343±.001 |
| $r_2$ | .05±.01 | .39±.01 | .31±.01 | .20±.03 | .130±.004 |
| $\varphi_k(r_2) - 2\varphi_k(r_1)$ | .46±.03 | -.163±.003 | .450±.007 | .20±.03 | .338±.004 |

The dependence of the semiamplitude of the main wave ($r_1$) of the light curve versus the minimal value of the spectral type (visual) is given in the fig. 3. The data for $r_1$ (visual) for 48 stars were used from the article by Kudashkina & Andronov (1996).

The periodogram analysis detects the several frequencies, which are multiple of main. The results are tabulated in the table 4. The example of the periodograms is shown on the fig. 4. *S(f)* – height of peak at the periodogram, which is a square of the correlation coefficient between the observations and the sine fit.

***Table 4. The results of the periodogram analysis.***

| **P, d** | **O Cet** | **R Aql** | **R Leo** | **RR Aql** | **S Car** | **$L_2$ Pup** | **U Her** | **S Ori** | **S Scl** | **X Oph** |
|---|---|---|---|---|---|---|---|---|---|---|
| **P(H)** | 333.52 | 280.01 | 312.79 | 390.89 | 149.94 | 137.14 | 407.00 | 414.33 | 367.03 | 332.56 |
| **P(K)** | 333.79 | 280.60 | 311.03 | 390.67 | 146.77 | 137.19 | 406.87 | 414.00 | 366.95 | 332.80 |
| **P(vis)** | 333.33 | 280.84 | 314.16 | 391.7 | 150.05 | 137.16 | 407.06 | 414.46 | 367.86 | 332.37 |
| **P/5** | | | | | | | | | 73.51 | |
| **P/3** | | | | 129.85 | | | 135.79 | 138.53 | 122.52 | |
| **P/2** | 166.62 | | 156.73 | 180.27 | 74.99 | | 203.56 | 206.58 | | 166.15 |
| **~P, ~1.5P** | 532.83 | | | | | 134.45 | | 394.5 | 649 | 316.8 |
| **2P** | | 582.76 | 626.43 | | | | | | | |
| **~3P** | | | | | 425.8 | | 1263 | | | |
| **~4P** | 1420 | | 1146 | 1579 | | | | | | |
| **>10P** | | 5461 | | | | 5511 | | | | |

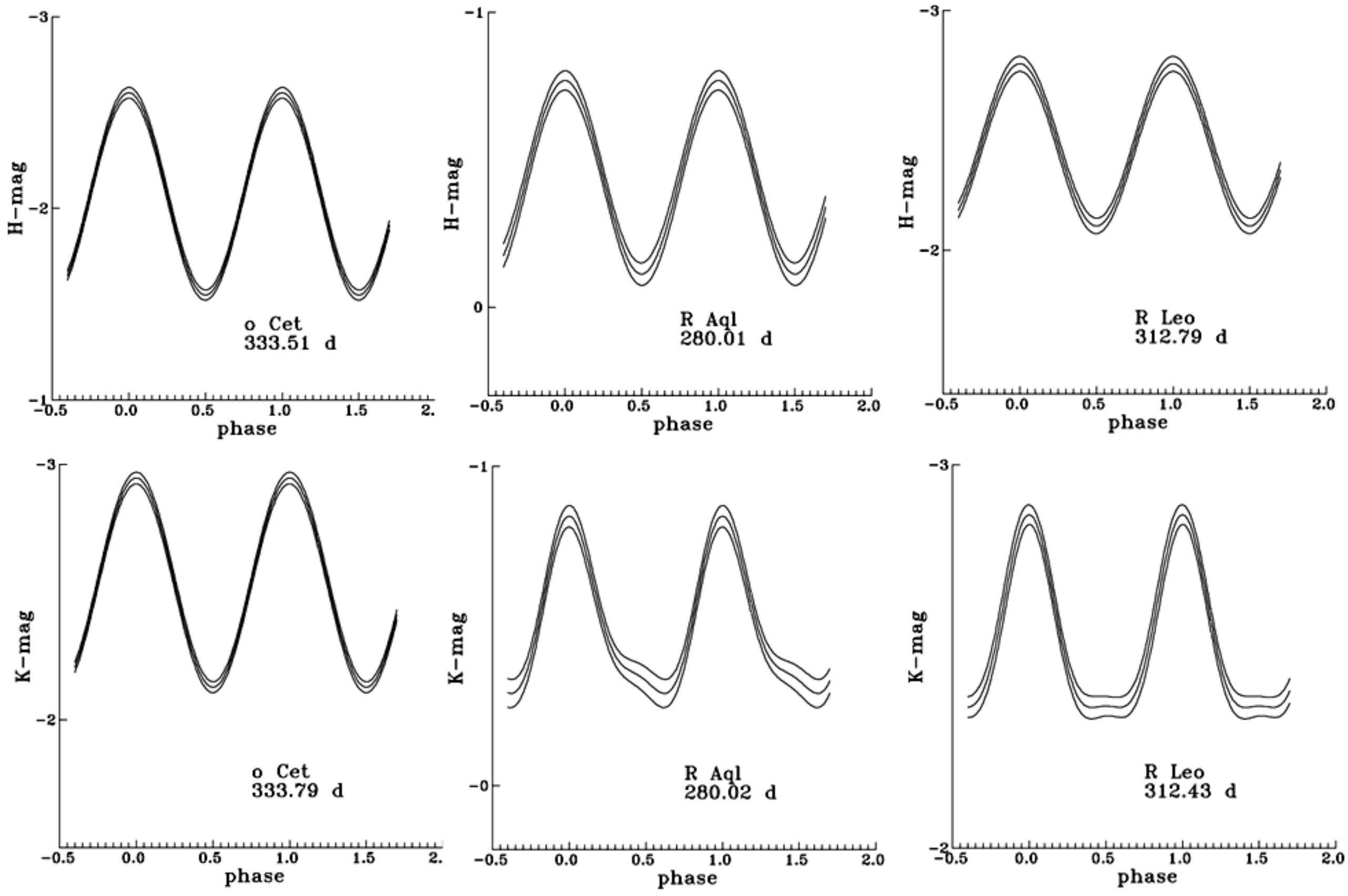

Fig. 1. The example of mean light curves of the Mira-type stars in H and K bands. There are fit line and ±1σ in the figure.

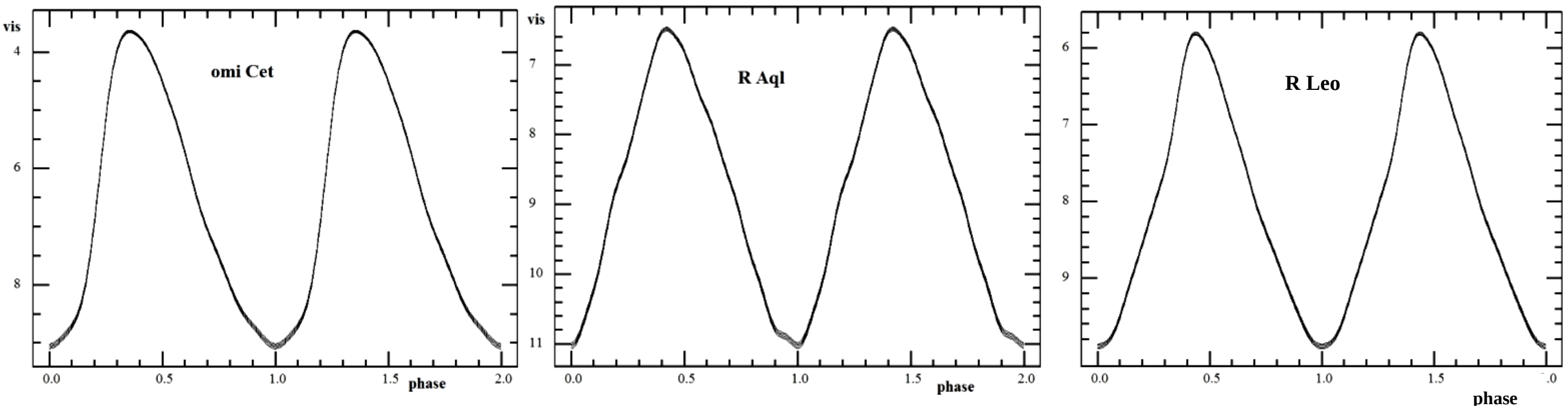

Fig. 2. The same for the visual light curves.

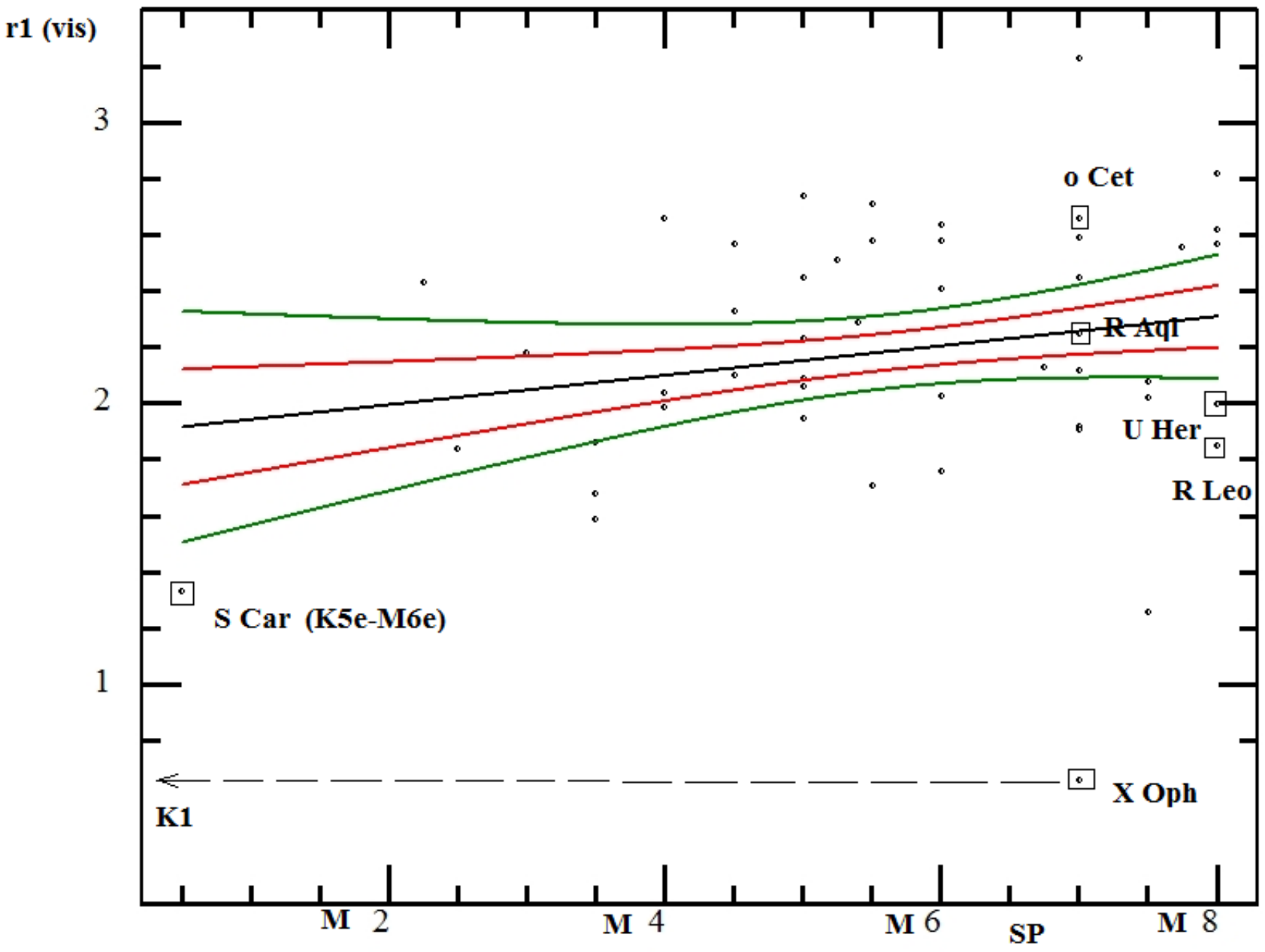

Fig. 3. The spectral types are from GCVS (the average values). The several individual stars are point. If for the star X Oph to take on the value from table 1, than the location of this star shifts along the pointer. The linear approximation have been performed using the program MCV (Andronov & Baklanov, 2004). The line regression coefficients are not statistically significant and are not listed here.

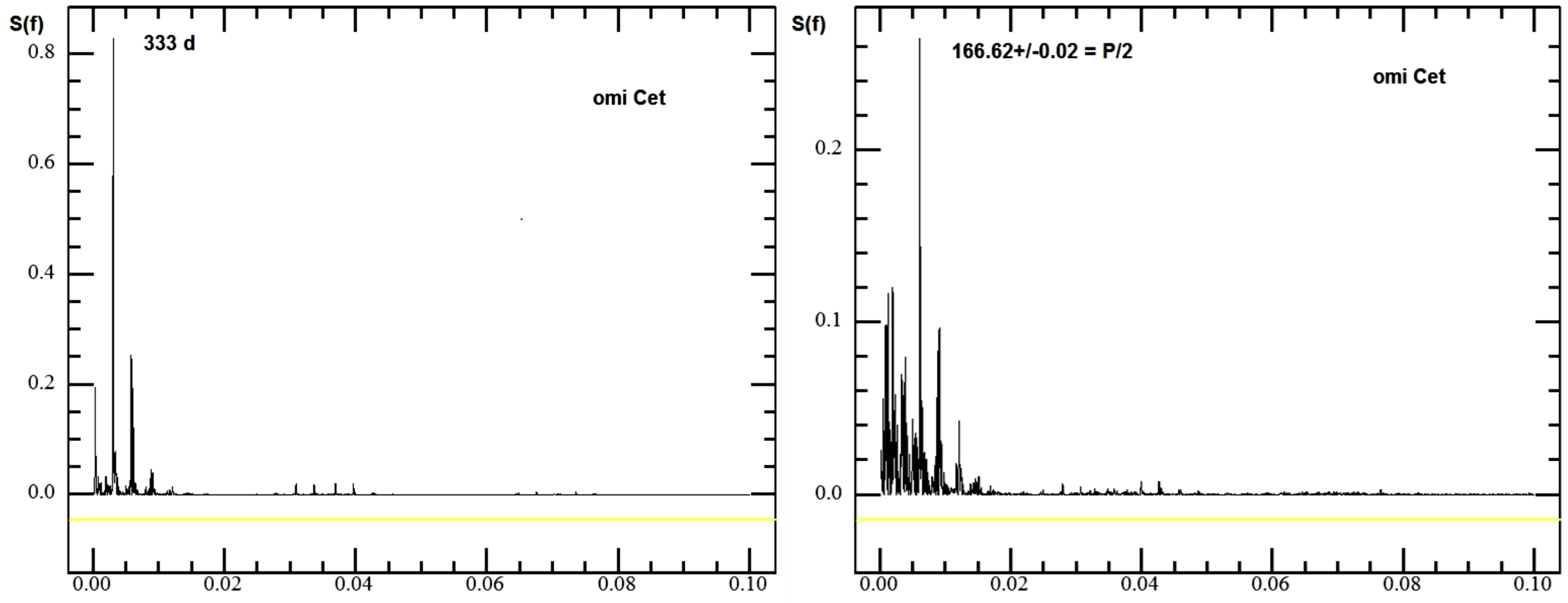

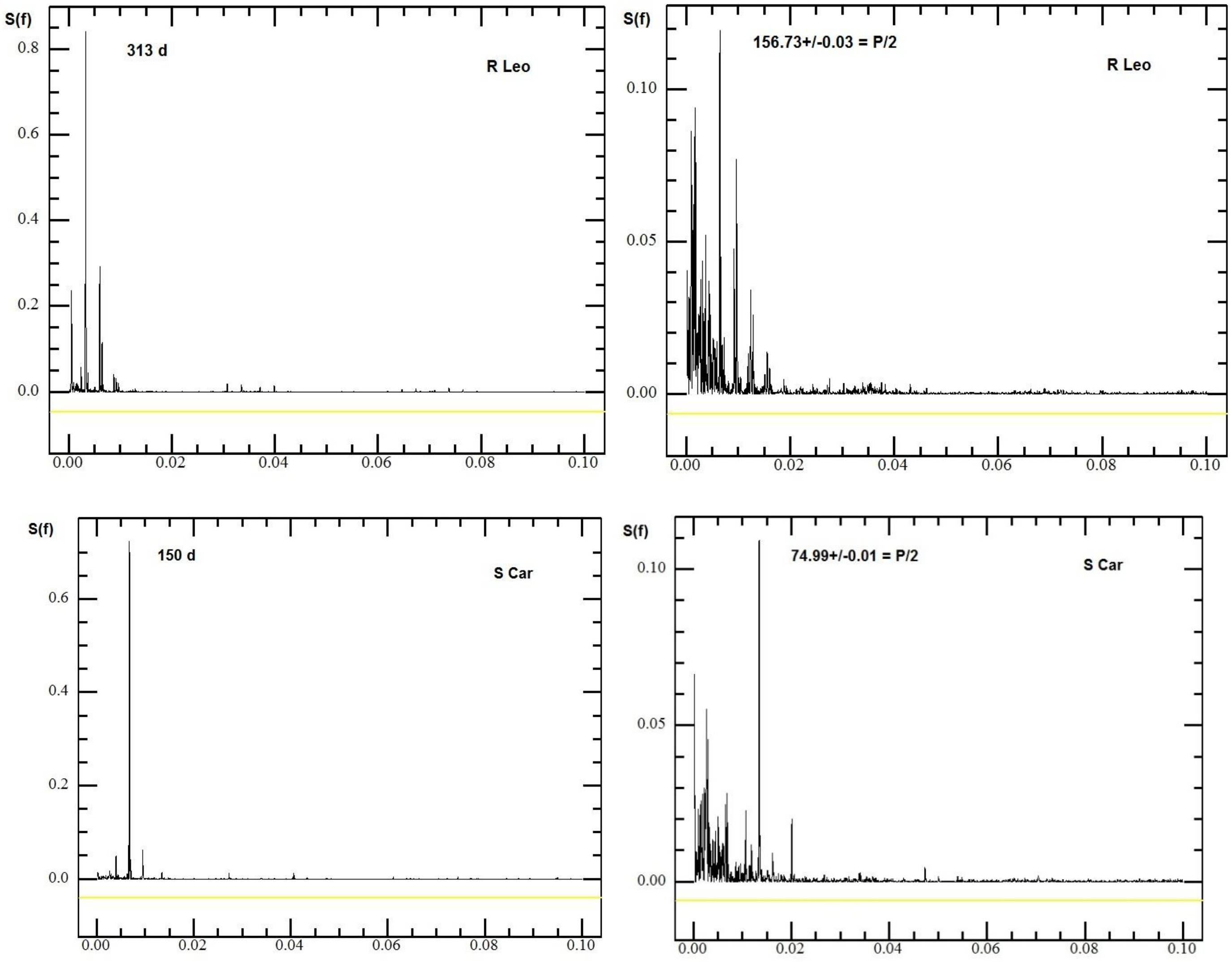

Fig. 4. The examples of the results of the periodogram analysis: test function S(f) vs frequency f=1/P.

References

1. Whitelock P.A., Marang F., Feast M. Mon. Not. R. Astron. Soc., 2000, 319, 728. (2000MNRAS.319..728W).
2. Andronov I.L., Odessa Astronomical Publications, 1994, v.7, 49. (1994OAP.....7...49A).
3. Kudashkina L.S., Andronov I.L. 1996: Odessa Astron. Publ., 9, 108. (1996OAP.....9..108K).
4. Kudashkina L.S., Andronov I.L., Proc. IAU Symp. № 180 "Planetary Nebulae", Groningen (The Netherlands). – 1996. – P.353.
5. Andronov I.L., 2003, ASP Conf. Ser. 292, 391. (2003ASPC..292..391A).
6. Andronov I.L., Baklanov A.V. Astronomical School's Report, 2004, 5, 264. (2004AstSR...5..264A )